# A situated agent-based model to reveal irrigators' options behind their actions under institutional arrangements in Southern France

Bastien Richard[1,2]*, Bruno Bonté[1], Olivier Barreteau[1], and Isabelle Braud[2]

[1]G-EAU, Univ Montpellier, AgroParisTech, BRGM, CIRAD, IRD, INRAE, Institut Agro, Montpellier, France
[2]INRAE, RiverLy, Villeurbanne, France

## Abstract

There has been little exploration of the explicit simulation of the set of options of actors in agent-based models and its evolution over time. This study proposes to use affordances as intermediate entities between agents' environment and agent actions. We illustrated the approach on a typical gravity-fed network in the South-East of France to explore how the abandonment of traditional sharing of water changes the irrigators' options to irrigate. We simulated a typical dry year irrigation season under two institutional arrangements (i.e. traditional coordination through daily slots and its abandonment). Simulation results are consistent with field surveys, and reveal an increase in the number of internal conflicts among irrigators as the counterpart of the abandonment of traditional sharing of water. They also highlight the consequences of the heterogeneity of the irrigators' interests within the collective institution. The sensitivity analysis of the model allowed identification of optimal modalities of coordination, and a potential compromise between past and current institutional arrangements. The key benefits of using affordances in ABM lie in the study of their population dynamics for characterizing the interaction situations between actors and their environment and for better understanding the model dynamics.

## Keywords

collective irrigation; agent-based model; situated action; affordance; institutional arrangement

## Code availability

The WatASit model was developed using the CORMAS modeling and simulation platform (Bousquet et al., 1998; Bommel et al., 2016). The model code and related data are publicly available at https://www.comses.net/codebases/0d8dcaf1-8772-4e57-9f03-1f6c062bbe60/releases/1.2.0/.

# 1. Introduction

Throughout history, local water management has been mainly carried out by collective institutions (Sanchis-Ibor and Molle, 2019). According to Loubier et al. (2019), among the 2000 networks managed by a collective institution in France, a quarter is gravity-fed. Most of them cover rather small irrigated areas and have a small number of members. They constitute complex Socio-Hydrological-Systems in which hydrological processes and human activities co-evolve and are linked through various interactions.

This is the case in the Buëch River basin (South East of France) where institutional arrangements in small gravity networks have recently undergone profound changes such as the abandonment of traditional coordination through daily slots (Richard et al., 2020). Such change in their institutional arrangements could affect their capacity to irrigate by modifying the operational functioning of the water network (Plusquellec, 1988; Malaterre, 2008). In particular, new inequalities between farmers in their ability to operate irrigation during the course of an irrigation campaign could appear (Richard et al., 2020).

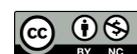

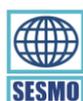

Agent-based models (ABMs) are commonly used for investigating common-pool resource management issues such as those at stake in Socio-Hydrological Systems. They are useful for representing a set of interacting agents that perform their tasks according to their specific objectives and available resources (Bousquet & Le Page, 2004; Gleizes et al., 2011). The design and analysis of ABM outputs usually involves estimating the effect of parameter values and external drivers on indicators, summarizing the time series of agents or environmental state variables through sensitivity analysis, spatio-temporal analysis, visualization, or communication to non-expert stakeholders (Lee et al., 2015). Explicitly simulating the set of options of agents and its evolution over time is still little explored and could give access to an intermediate level of outcomes useful to better understand such indicators. This could be particularly interesting in situations that are difficult to predict in advance, such as farmers' short-term adaptations to a limiting water resource (Reynaud, 2009), when each farmer's operations are likely to quickly affect the capabilities of others (Daydé et al., 2014).

In the vein of the Theory of Situated Action (Dreyfus, 1972; Suchman, 1987), several authors have proposed introducing the concept of affordance in ABM to focus on the concrete modalities of the implementation of action in a given situation, i.e. at a given place and time (Cornwell et al., 2003; Sequeira et al., 2007; Papasimeon, 2009; Afoutni et al., 2014). In such situated ABMs, the actor's behavior is determined by a set of possible actions called affordances (Gibson, 1977) and continuously changing according to the current situation, forcing the actor to adapt to it (Gibson, 1977).

We have developed the situated WatASit ABM simulating the operations of irrigators sharing water through a common network during an irrigation campaign. The objective of the study is to show how the use of affordances as intermediate objects between the environment and agents of the ABM allows the irrigator agents' options to irrigate to be made explicit, providing new kinds of outputs. The approach is illustrated using a typical gravity network in the South-East of France and is used to explore how irrigators' options change under two institutional arrangements for the sharing of water (i.e. traditional network coordination through daily slots and its abandonment).

The next section presents the study site (Section 2.1), the data collection (Section 2.2), the use of the affordance concept to represent irrigators' options to irrigate (Section 2.3), and an overview of the WatASit ABM (Section 2.4). Section 3 describes the simulation results, and Section 4 discusses them before the concluding remarks.

## 2. Materials and Methods

### 2.1 Study site

Our case study was conducted in the Buëch River basin which extends over an area of 1,490 km$^2$, mainly within the Hautes-Alpes County (Figure 1). The Buëch River basin is a typical medium mountain basin in the South-East of France with water management facing operational evolutions to comply with new environmental considerations and associated norms. Collective irrigation is the most important use of water resources during the low-flow period. Due to the quantitative gap, the river basin institution has adopted a water resources management plan to reduce the water abstraction from irrigation. This plan includes the modernization of gravity networks to finely pilot water withdrawals, to better regulate water flows, and to reduce network seepage.

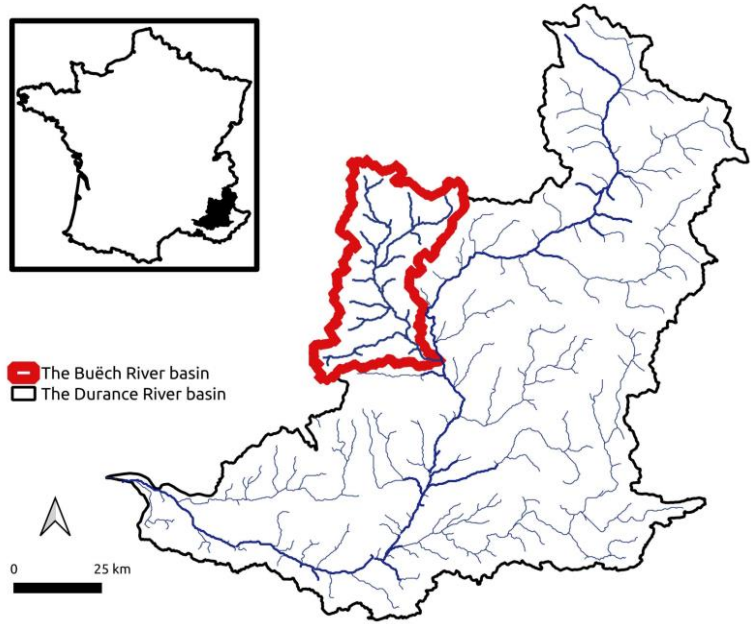

**Figure 1:** Location of the Buëch River basin (red contour line) in France. The Buëch River is a tributary of the Durance River basin (black contour line).

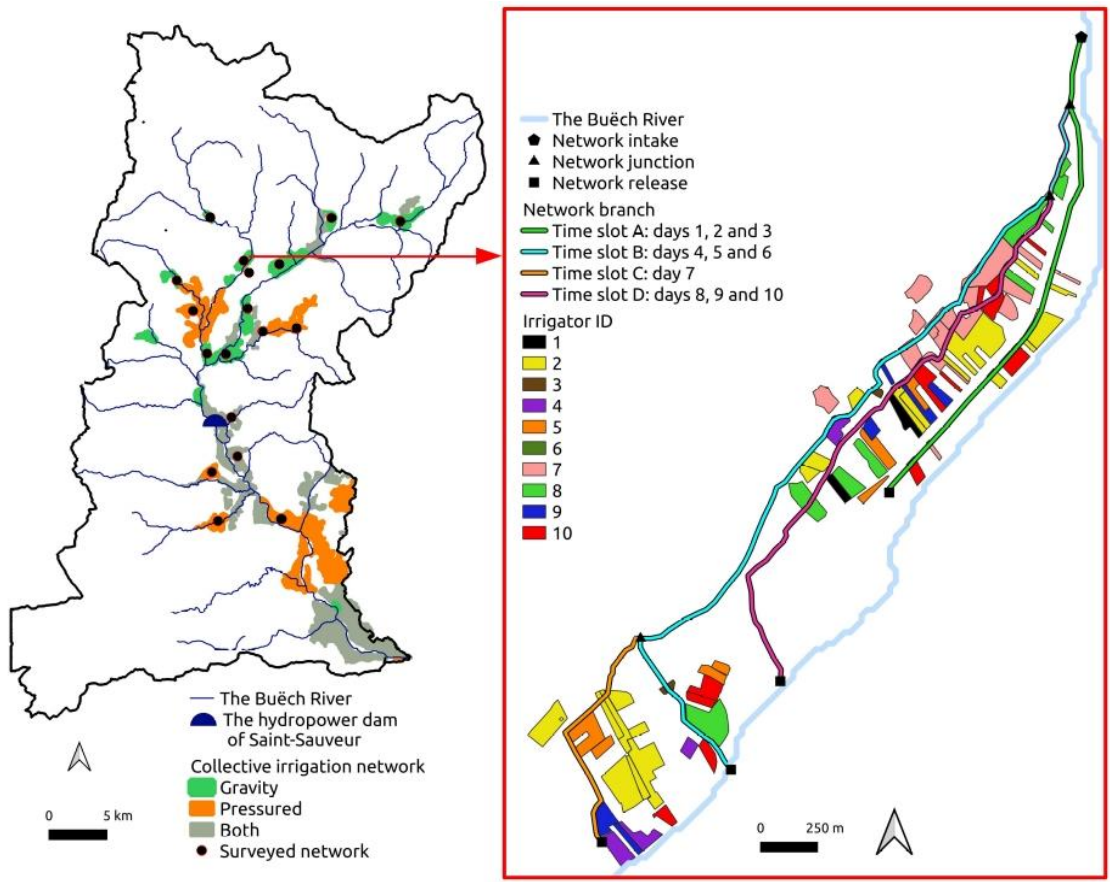

**Figure 2:** The collective irrigation networks of the Buëch River basin (left-side), and the study area (red box). In the red box are mapped the irrigated plots served with water by the Aspres-Sur-Buëch gravity network (source: RPG 2017 and BD Hydra consulted in March 2019). The colors of the plots correspond to the irrigators' identifiers. When the network is coordinated through daily slots, each colored line designates the branch that is watered during a given time slot. For example, the green branch flows during slot A, i.e. during the first 3 days of each 10-days period. The blue branch flows during the next 3 days, the orange branch during day 7, and the pink branch during the last 3 days, then again the green branch is put in water.

Figure 2 maps the collective irrigation networks within the Buëch River basin. Gravity networks (green areas) are mainly located upstream of the hydropower dam of Saint-Sauveur, with unsecured water access during the low-flow period. After having investigated the changes at work in irrigation management practices of several collective networks of the Buëch basin, we observed an ongoing individualization of the operational management of irrigation in the gravity networks (Richard, 2020).

We have chosen to zoom in on the Aspres-Sur-Buëch gravity network as a case study because it is fairly representative of the gravity networks of the Buëch River basin in terms of location (in the upstream part of the basin), irrigable surface area (with 75 ha close to the average of 50 ha in the Buëch in 2017), and crop types (with 38.9 % meadows, 29.5 % fodders and 23.7 % cereals in 2017). It is also a striking example of change in operational management with the abandonment of traditional coordination of the network through daily slots progressively over the past 15 years. The study area includes 77 irrigated plots shared between 10 irrigators (Figure 2: red box).

Water sharing among the Aspres-Sur-Buëch irrigators through a daily slot calendar has been gradually abandoned during the last 15 years as it was temporally very restrictive according to the interviewees. We have captured the latest version in place, the one known to all the interviewees. The branches of the canal were watered according to 4 time slots (A, B, C, and D, Figure 2). Under the water sharing arrangement, during each time slot, water flows only in the corresponding branches designated by the calendar (e.g. the green branch during slot A, the blue branch during slot B, etc.). The different daily slots follow one after the other over 10 days. Currently, irrigators do not coordinate the water network: the water flows simultaneously and continuously in all the branches of the canal until the end of the irrigation campaign.

## 2.2 Data collection

To capture the functioning of the water network to operate irrigation, we conducted field surveys on the case study as well as direct observation of irrigators' practices and semi-structured interviews about the irrigation campaign that took place between May and September 2017. The interviews were conducted with farmers 2, 7, 9, and 10 because of the time constraints of other farmers during this period. We also interviewed the technician in charge of the water network regulation, and since farmer 9 was also the President of the irrigator union, the information collected covered the entire irrigated command area. Key information collected is further detailed in the Supplementary Material.

## 2.3 Affordances to represent irrigators' options to irrigate

In the vein of the Theory of Situated Action (Dreyfus, 1972; Suchman, 1987), Gibson's ecological approach refers to affordance as any possibility of interaction offered to actors by their environment: "the affordances of the environment are what it offers the animal, what it provides or furnishes, either for good or ill" (Gibson, 1977). Gibson argues that species live in an environment designed as a set of affordances complementary to them (Simone, 2011). Man is the clearest example of this since he produces new affordances to make his environment easier to live in. Thus, actors use the possible actions (i.e. affordances) offered to them by the other elements (actors or objects) they perceived (Gibson, 1986; Johnston & Brennan, 1996; Clancey, 2002). So the action is performed from a limited subset of possible actions occurring in a precise place at a given time. Figure 3 represents the actions performed within a set of available options. Each grey or black spot is figuring one available option of action: 3 options in the first time step, only one in the second, 5 in the third, none in the fourth and 3 in the last time step - none of which is chosen. Variations in the concept of affordance have been proposed (Turvey, 1992; Chemero, 2003; Stoeffregen, 2003), but they have in common that:

- An affordance is action-oriented: the environment influences the actors' actions,
- An affordance is significant: the perception of an object is the perception of what it is possible to do with it in terms of action,
- An affordance is directly perceived: the actor does not need to have a symbolic representation of its environment to understand it and act accordingly.

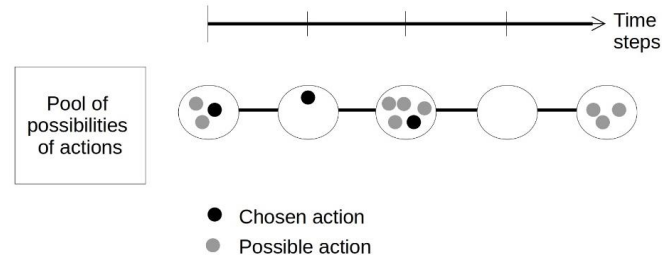

**Figure 3:** Schematic example of affordances as possible actions (grey spots) constraining the actor's choice to act (black spots) in a given situation (i.e. at a given place and time).

In our modeling approach, each action and possible actions of Figure 3 is explicitly represented. The actions are represented by models of actions and possible actions are represented by affordances which are both modeled as situated entities of the model. Affordances are perceived by the agents present in the model and when selected, they trigger an action involving the actor and an object of the environment. Actions and affordances have their own dynamics of appearance and disappearance coordinated with other dynamics of the model by the simulation scheduler. The behaviors of the agents represented in the model is then partly coded as the dynamics of a population of affordances that the agents can perceive and as the actions that agents perform. As it is described in the next section, we will say that an agent that perform an action is an "actuator" of this action and that an entity of the environment that is affected by an action is a "passive object". In our model, all agents are actuators and all entities of the environment that may be affected by agents' actions are passive objects. In addition to forcing the modeler to explicitly describe the conditions in which options are made available to actors, it eases the analyses of actors' options by studying the population dynamics of affordances in the model. In the following we will use the term "option" when we speak about the options of farmers in the real world, and the term "affordance" when we refer to their representation in the model.

Several authors have pointed out the concept of affordance to better understand the principle « structure influences behavior » (Forrester, 1968; Harris, 1990). Indeed, the affordances describe what behaviors are possible:

$$structure \rightarrow affordance \rightarrow behavior$$

What happens when we change the structure of a system? A change in structure propagates to a change in the affordances, which might propagate to changes in behavior:

$$\Delta structure \rightarrow \Delta affordance \rightarrow \Delta behavior$$

The objective of using affordances as intermediate entities between system structure and agents' actions is to analyze the dynamics of the affordances population during the simulations. This analysis is meant to be used to discuss the effect of a modification of the system structure (such as a change in the institutional arrangement for the sharing of water) on actors' behaviors in the light of the options offered to them with this new structure.

In our modeling approach, we represent two structures of the system by implementing two configurations of our model, representing two institutional arrangements (later called "DailySlots" and "NoSlots"). We consider that the study of the population dynamics of affordances might be a meaningful output to relate more standard model outputs (number of plots abandon, heterogeneity of irrigation, etc.) with those two model configurations.

## 2.4 Model description

We use the ODD protocol (Grimm et al., 2006; 2010) to describe the model.

### 2.4.1 Overview

The WatASit ABM is designed to simulate the irrigation operations of irrigators sharing a common water network during a collective irrigation campaign. It explicitly represents the irrigation options left by the operational constraints of the irrigators. The constraints taken into account in WatASit are presented in Table 1. In WatASit, an irrigation possibility is generated on the plots where these constraints make irrigation possible at a given hourly time step.

**Table 1:** The network-specific constraints in the WatASit model.

| Network constraint | Case study specification |
|---|---|
| Number of irrigators per farm | *One irrigator per farm* |
| Number of simultaneous irrigations per irrigator | *One irrigation at a time per irrigator* |
| Daily time window (maximum daily working time of each irrigator) | *12h* |
| Plot flood duration | *Fixed* |
| Target irrigation dose | *Fixed* |
| Required branch canal flow serving the plot floodgate | *≥ plot flood rate (Qflood)* |
| Functioning of the network while raining | *Irrigation is not triggered if there is precipitation* |

***Entities***

The model is based on the distinction between the elements that are involved in irrigation operations (called operational entities) and the areas over which operational entities can operate (called spatial entities) which are the farm plot, the farm, and the irrigation scheme area.

We distinguished two types of operational entities, the actuators, and the passive objects. An actuator is an entity with capacities to carry out actions with the passive objects located on the spatial entities under its control. In the model, the actuators are the irrigator agents. A passive object is an entity such as a floodgate that is actionable by an actuator to act. In the model, the passive objects considered are the network branch, intake, junction, and release points, and the floodgate at each farm plot, the crops, and the irrigators.

The third kind of entity is artifacts. WatASit considers artifacts to make explicit some abstract things of the real world such as options to irrigate (represented as affordances) and actions. Affordances result from the interactions between an actuator (i.e. an irrigator agent) and a passive object (i.e. a piece of irrigation equipment represented in the model). The decision process of the actuators consists of choosing among the affordances it perceives. On a given farm, they can interact according to conditions that define the situation of interaction. If conditions are fulfilled, an irrigation affordance may be generated. Whenever it is selected, it becomes an irrigation action. As we need a neutral and abstract level to detect and reify these artifacts within each farm entity, we designed dedicated entities, called situation controllers which are responsible for the coordination of Actions and Affordances at the farm level. Inspired by Afoutni et al. (2014) "place-agents", each farm is associated with a situation controller.

The types of affordances considered in our model are described in Table 2. In particular, the *AskMoreWater* affordance is generated when the water level in the branch of a canal is not sufficient with regard to the flow required to irrigate a farm plot. It results from a temporary drop in the water level due to the withdrawal by an irrigator to the detriment of another irrigator. We consider such a situation as a potential conflict for water access at this moment between these irrigators, and that the presence of an *AskMoreWater* affordance is revealing such potential conflict in the simulation. All entities in the model, their characteristics, typical values, and data sources are available in the Supplementary Material, as well as a class diagram describing the structure of the model.

***Process overview and scheduling***

The model is based on a double-time step. Each day, there is first an initialization of the current precipitation conditions, and also an update of the number of days since the crops have not been irrigated. Then, every hour, the flow is updated in the network according to network junction state (i.e. "opened" or "closed") and ended actions. Irrigation affordances and actions are then generated on each farm by the associated situation controllers according the list of actuators and passive objects present in the farm and to their states. Depending on the irrigator's decision-making (see Section 2.4.3), an affordance can be chosen to make a flood action or ask for more water in the canal. An activity diagram describing process scheduling is available in the Supplementary Material.

**Table 2:** The affordances considered in WatASit deployed on the Aspres-Sur-Buëch case study. Associated actuator/passive object pairings for each affordance, generation conditions as well as the execution conditions and effect of the corresponding action are presented. Q is the flow of the passive object, Qmin and Qmax are the minimum and maximum flow, respectively. Qintake is the flow entering into the network, and Qrung is the flow increase when an *AskMoreWater* affordance is performed. All variables are detailed in supplementary materials.

| Affordance name | Actuator / passive object | Affordance generation conditions | Action execution conditions | Action execution effect |
|---|---|---|---|---|
| *Flood* | Irrigator/plot floodgate | *Actuator* availability No precipitation *Passive object* status $(Q \geq Qmin)$ | *Passive object* status $(Q - Qmin \geq 0)$ Action duration $< D$ | $Q = Q - Qmin$ |
| *AskMoreWater* | Irrigator/plot floodgate | *Actuator* availability No precipitation *Passive object* status $(Q < Qmin)$ | None | $Qintake = \min(Qmax, Qintake + Qrung)$ |
| *DoSomethingElse* | Irrigator/another passive object | *Actuator* availability | None | None |

## 2.4.2 Design concepts

According to the theory of Situated Action (Dreyfus, 1972, Suchman, 1987), WatASit represents the phase of implementation of actions according to the operational constraints of the actors. In the model, the behavior of the irrigator agents is determined by their affordances to irrigate, which are re-evaluated at each hourly time step. Each agent chooses none or one affordance among the affordances he perceives and performs it.

Stochasticity can be included for climate forcing, but the initialization of the cropping systems for each irrigator is generally made from a geo-referenced dataset of the irrigation campaign year and the model is usually deterministic. Agents do not predict the consequences of their behavior and do not change their strategy as a consequence of their behavior. However, the dynamics of agents' affordances, which are part of the description of agents' behavior, depend on the context of the moment, forcing them to modify their behavior accordingly.

WatASit allows direct interactions between the actuators (e.g. the irrigator agents) and the passive objects of their spatial environment. When one actuator has performed an affordance, all affordances in the environment are updated. Irrigator agents interact indirectly with each other by reducing the amount of water available in the network, which affects the affordances of the other agents sharing the same or a downstream branch.

Model simulation outputs are collected by recording attributes of model entities directly in the R Software through the *Rcoupler* application programming interface that has been developed in this work to configure and pilot simulations of the WatASit model. The final outputs of the model are the location and number of plots that were abandoned for irrigation at the end of the simulated irrigation campaign between May 1[st] and September 30[th]. The irrigators' options explicitly simulated each hour through the affordances artifacts (Table 2) constitute an additional output of the model that helps explaining the final outputs. They allow counting the cumulative number of days when irrigators may choose to flood one of their plots (by monitoring *Flood* affordances) or when they can't due to local water shortage (by monitoring *AskMoreWater* affordances).

## 2.4.3 Details

The model has been already deployed to represent the Aspres-Sur-Buëch case study (Richard et al., 2020) for which the specific parameterization has been detailed. The simulation period covers the irrigation campaign from May 1st to September 30th, with the year 2017 taken as a reference and representative of a dry year. Spatial entities were initialized using a pre-processing that consists of rasterizing the farm plot shapefiles (Table 3) onto a 54 x 44 cell grid with a resolution of 75 m. The water network was also initialized using shapefiles (Table 3). At the initialization, farm plots served by the network were listed for each canal branch. Daily precipitation input comes from the French near-surface SAFRAN reanalysis (Vidal et al., 2010).

**Table 3:** Datatypes and data sources.

| Data type | Data source |
|---|---|
| Daily precipitation | SAFRAN reanalysis (Vidal et al., 2010) |
| Farm plot shapefiles, areas and crop types | "*Registre Parcellaire Graphique*" (RPG) 2017[1] |
| Water network shapefile | BD HYDRA (v2) 2015 |

Decision-making of the irrigator agents consists of irrigating crops after several days without sufficient precipitation inputs to reduce the number of successive non-irrigated days on their plots (Figure 4). As water is not always sufficiently available to irrigate a plot, irrigator agents have three kinds of affordances: irrigate a plot when a *Flood* affordance is available, ask for an increase in water flow in the network when an *AskMoreWater* affordance is available, and do something else during this time step. Finally, a plot for which the irrigator agent has not had any Flood affordance for a certain period of time is abandoned for irrigation (see Figure 4).

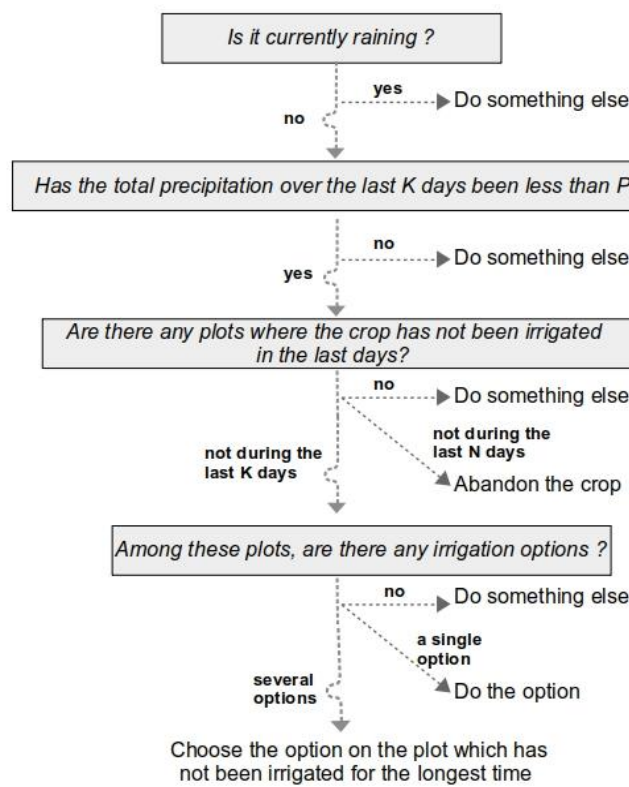

**Figure 4:** Operational decision-making of the irrigator agents. Typical values are K=12 days, P=120 mm, and N = 30 days

According to field surveys, irrigators have abandoned the network coordination through daily slots and were able to trigger irrigation without coordination during the 2017 irrigation campaign. In this study, we first initialize the model and simulate the 2017 irrigation campaign considering the network flow is coordinated ("DailySlots" configuration of the model). Then, we initialize the model without coordination of the network by the irrigators ("NoSlots" configuration) and simulate again the 2017 campaign.

### 2.4.4 Sub-models

Affordance generation sub-model, action execution sub-model, and simplified hydraulic sub-model of the water network are fully described in the Supplementary Material.

---

1      A version 2.0 distributed since 2015 is directly accessible online at http://professionnels.ign.fr/rpg

# 3. Simulation results

## 3.1 Final outcomes of the model simulations: abandoned plots for irrigation

In this section, we compare the plots where irrigation was abandoned by the irrigator agents at the end of the simulated irrigation campaign under the two institutional arrangements (i.e. DailySlots and NoSlots). In the NoSlots configuration, 23 plots were abandoned for irrigation instead of 3 in the DailySlots configuration (Figure 5). Irrigator agent 2 is the only one who abandoned irrigation in one or more plots under the DailySlots model configurations. Irrigator agent 4 is the most impacted by the abandonment of the coordination through daily slots with 3 of his 4 plots abandoned (i.e. 75% of abandons) in the NoSlots configuration. Note that all the abandoned plots for irrigation are located in the downstream part of the irrigation network, making a strong dichotomy with the upstream part. This is consistent with the observations made during field surveys: all downstream plots were abandoned for irrigation when irrigators did not coordinate the irrigation network during the 2017 campaign.

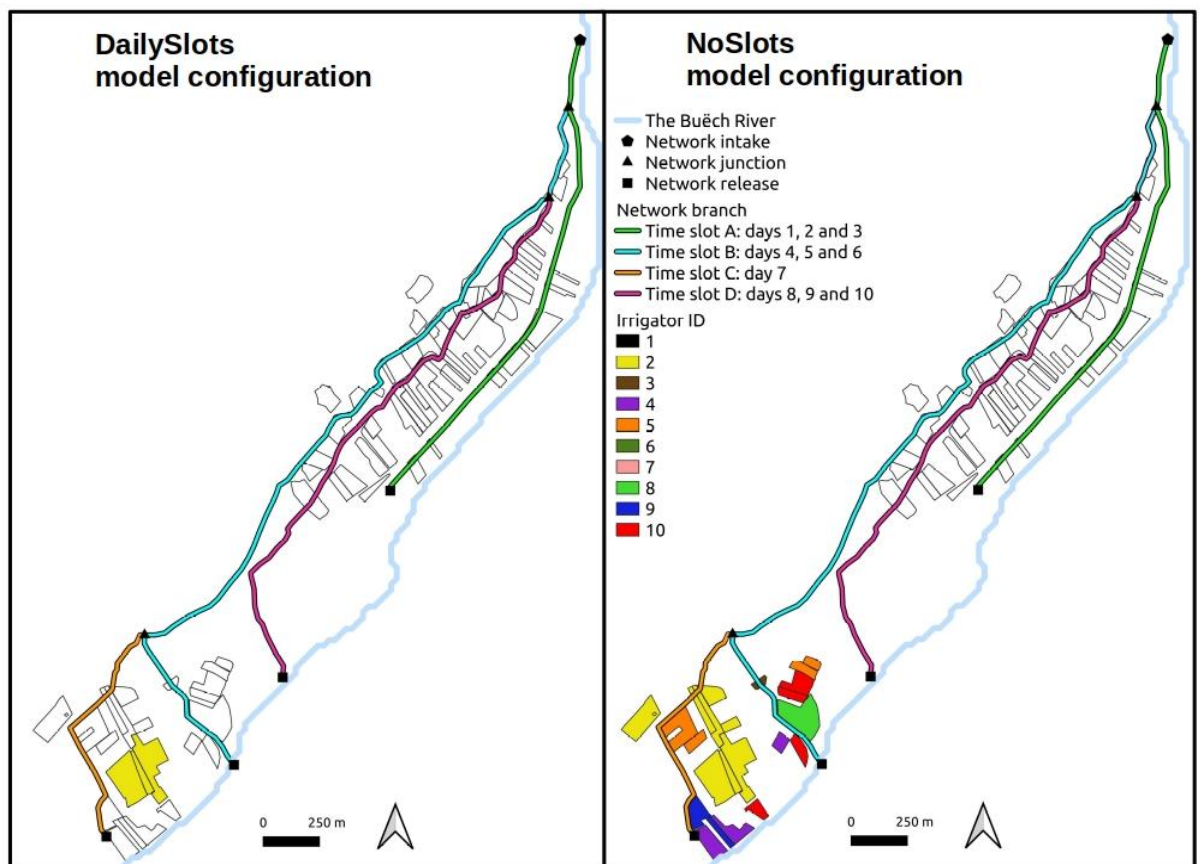

**Figure 5:** Colors denote abandoned plots for irrigation of irrigator agents 1 to 10 for the DailySlots (left-side) and NoSlots (right-side) model configurations at the end of the simulated irrigation campaign (May 1st – September 30th).

## 3.2 Intermediate outcomes of the model simulations: affordances

In this section, we start by giving an overview of the affordances simulated by the model, and then we zoom in on two individual pathways with high impact.

### 3.2.1 Spatial distribution of irrigator agents' affordances

The WatASit model simulates the dynamics of the population of Flood and AskMoreWater affordances (see Table 2). A Flood affordance generated by the model means that the canal branch supplying the plot is flowing enough to trigger gravity irrigation, while the generation of an AskMoreWater affordance means that the canal branch is not flowing enough at the time the irrigator agent wants to irrigate it. Figure 6 maps, for each plot, the cumulated number of days of the presence of a Flood affordance (blue color classes) and a AskMoreWater (red color classes) affordance during the simulated irrigation campaign from May 1st to September 30th.

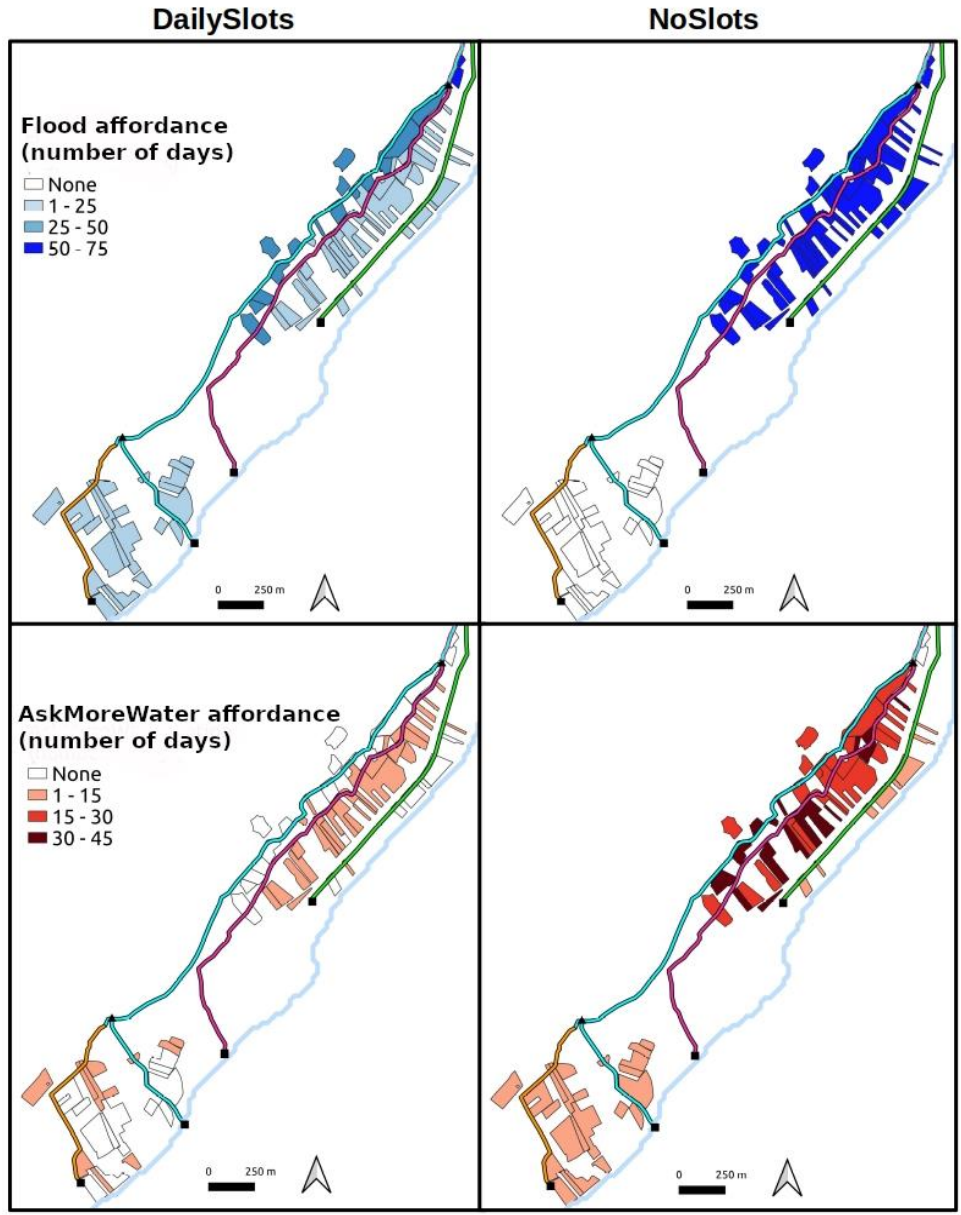

**Figure 6:** Spatial distribution of the cumulative number of days the Flood affordances (blue colors) and AskMoreWater affordances (red colors) were present under the DailySlots and NoSlots model configurations, during the simulation period from May 1st to September 30th. Colored lines denote the time slots of the canal branches in the DailySlots model configuration: days 1, 2, and 3 for the green branch (slot A), days 4, 5, and 6 for the pink branch (slot B), day 7 for the cyan branch (slot C) and days 8, 9 and 10 for the orange branch (slot D).

In the DailySlots model configuration, Flood affordances were generated on all plots (blue colors) (Figure 6). The cumulative number of days of the presence of a Flood affordance during the irrigation campaign varies according to the length of time the canal branch serving a plot was flowing. The two plots located upstream (dark blue plots) are those with a cumulative number of days of presence greater than 50 because the canal branch that serves them with water (i.e. the blue one) flows during the B but also C and D slots (see Figure 5), i.e. 7 days out of 10. The plots having a cumulative number of days of presence between 25 and 50 (blue plots) are served during two slots (B and C), that is 4 days out of 10. Finally, plots with less than 25 days with at least one Flood affordance (light blue plots) are served during only one slot (A, B, C, or D), i.e. a maximum of 3 days.

In the NoSlots model configuration, no Flood affordance were generated in the plots located downstream of the network (white plots, see also Table 4). On the contrary, all the plots located upstream had a cumulative number of days of the presence of a Flood affordance higher than 50 (dark blue plots) with an average of 71 (Table 4). When irrigator agents don't coordinate the network (NoSlots), water flows simultaneously in all canal branches. This results in an upstream-downstream gradient of water flow, which splits at each branch junction, and is reduced by irrigation withdrawn and due to infiltration losses along canal branches. A dichotomy appears between the upstream and the downstream plots where the simulated flow is not sufficient to trigger any flood irrigation during the irrigation campaign, which explains their abandonment by the irrigator agents (see Figure 5 right-side). This is consistent with the words of the irrigators interviewed who say they prefer to focus on the upstream part of the irrigation scheme by having more continuous access to water rather than partial access over the entire area.

**Table 4:** Average cumulative number of days of the presence of a Flood affordance or a AskMoreWater affordance for all, upstream and downstream plots in the DailySlots and NoSlots model configurations at the end of the simulated irrigation campaign (May 1$^{st}$ – September 30$^{th}$).

|  | Flood affordances | | AskMoreWater affordances | |
|---|---|---|---|---|
|  | *DailySlots* | *NoSlots* | *DailySlots* | *NoSlots* |
| All plots | 21.42 | 49.79 | 3.22 | 22.01 |
| Upstream plots | 25.22 | 71.00 | 3.87 | 25.42 |
| Downstream plots | 12.48 | 0 | 1.69 | 14.00 |

The cumulative number of days of the presence of a AskMoreWater affordance (Figure 6, red color classes) is higher in the NoSlots model configuration than in the DailySlots one (6.8 times higher, Table 4), meaning plots are not served enough with water at the time the irrigator agent wants to flood them. This is not an intuitive result, in particular in the upstream plots concentrating a greater cumulative number of days of the presence of a Flood affordance. This shows that the presence of a Flood affordance in a greater number of upstream plots is not sufficient to avoid conflicts between irrigators for the sharing of water in the absence of network coordination.

This could explain why irrigators mention that although they have chosen to abandon traditional coordination by daily slots, they are not satisfied with the absence of coordination and plan the conversion to pressurized irrigation networks that allow on-demand irrigation that is less dependent on coordination between irrigators.

### 3.2.2 Two contrasting individual pathways

Irrigator agent 4 is the most impacted by the change in institutional arrangement. Figure 7 compares his trajectories in the two model configurations until it abandons irrigation. The percentage of plots with a Flood affordance reaches a maximum of 75 % when the network is coordinated through daily slots (Figure 7A black bars), instead of 25% in the absence of network coordination (green bars). Visualizing affordances that this agent perceives on each plot (Figure 7B) shows that he first irrigated the only plot with a Flood affordance (top line, black box, DOY 9). He then asked for more water to flood one of his downstream plots (second line, grey boxes, DOYS 9 to 30), before deciding to abandon them (red boxes) and flood his upstream plot again (top line, black box, DOY 38). The presence of AskMoreWater affordances (grey boxes) materializes insufficient water flow to trigger flood irrigation in the downstream branches of the canal when the network is not coordinated (NoSlots). It brings to a lack of Flood affordances at their location which causes their abandonment for irrigation. The abandonment of traditional sharing of water has reduced the number of his Flood affordances in the downstream part of the irrigation scheme, and limits, by the way, his spatial capacity to irrigate.

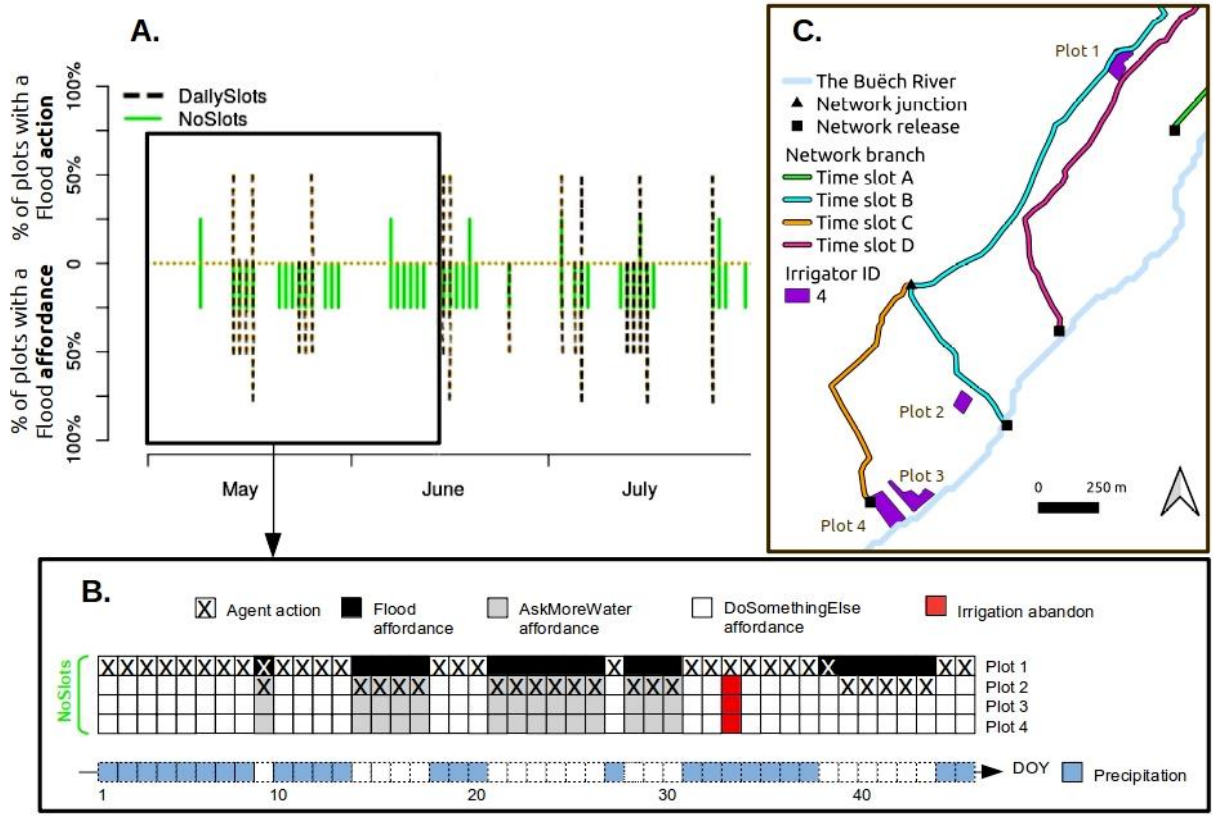

**Figure 7:** A) Bar chart of the percentage of plots of irrigator agent 4 with a Flood action (top) or a Flood affordance (bottom) over the simulated irrigation campaign from May 1st to September 30th in both DailySlots (dotted black bars) and NoSlots (green bars) model configurations. B) Visualization of irrigator agent 4's daily affordances in the NoSlots model configuration, from May 1st to June 14th. DOY is Day Of Year. C) Plot location of irrigator agent 4 within the network command area.

Irrigator agent 2 is the only one with abandoned plots when the network is coordinated through daily slots (Figure 5, yellow plots). Figure 8 looks at the trajectory in the DailySlots configuration of the model until he abandons irrigation on 3 of his 20 plots. In this case, its percentage of plots with a Flood affordance reaches a maximum of 25 % when the network is coordinated through daily slots (Figure 8A black bars), instead of more than 50% in the absence of network coordination (green bars). The absence of AskMoreWater affordances (Figure 8B) means that the water flow is always sufficient to trigger irrigation when the network is coordinated through daily slots. Plots 13 to 20 are served by the same canal branch (Figure 8C, orange branch). As irrigator agent 2 gets several Flood affordances during the same day (e.g. DOY 17), he irrigates as much as possible within the limit of the maximum daily working time (and according to the time required for each plot which depends on the area), but they are too numerous to be undertaken during time slot C. With the traditional institutional arrangement, the temporal constraint is predominant: irrigator agent 2 does not have time to irrigate plots 17 to 20 (DOY 17) which leads to them being abandoned for irrigation (DOY 33).

We better understand why irrigator 2 said he has an interest in abandoning traditional sharing of water to focus on his upstream plots and free up leeways to irrigate (as observed in the NoSlots configuration with an increase in the number of Flood affordances). On the contrary, irrigator 4 has no interest in abandoning traditional sharing of water as it made it possible to ensure a minimum flow downstream where three-quarters of his plots are located. This paradoxical position may have several explanations (e.g. misunderstanding of network functioning, political weakness, etc.), calling for further social investigations to understand these institutional dynamics.

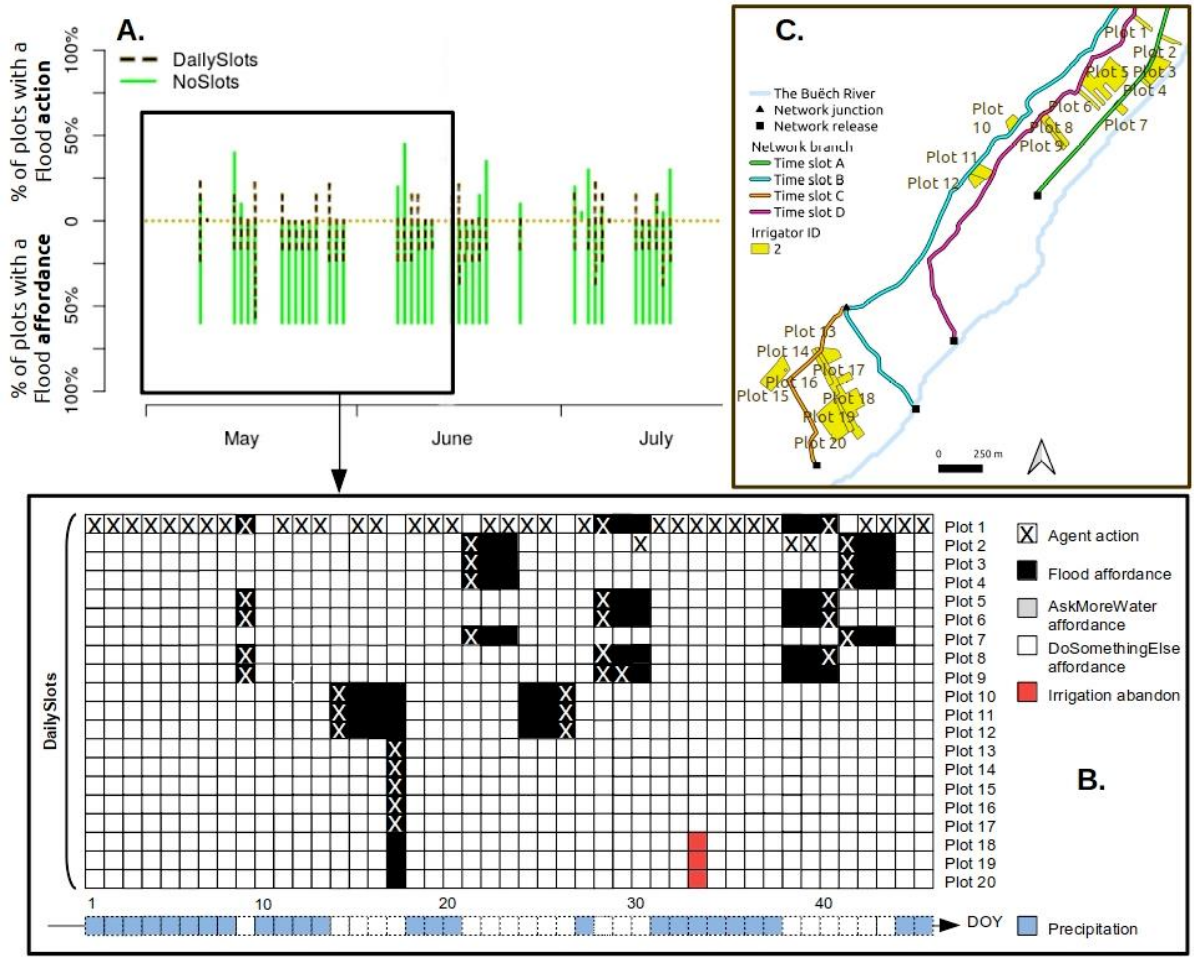

**Figure 8:** A) Bar chart of the percentage of plots of irrigator agent 2 with a Flood action (top) and a Flood affordance (bottom) over the simulated irrigation campaign from May 1st to September 30th in both DailySlots (dotted black bars) and NoSlots (green bars) model configurations. B) Visualization of irrigator agent 2's daily options in the DailySlots model configuration, from May 1st to June 14th. DOY is Day Of Year. C) Plot location of irrigator agent 2 within the network command area.

## 4. Discussion

### 4.1 Sensitivity of model outcomes to key forcing and parameters

We performed a one-at-a-time exploration to assess the influence of key model forcing (i.e. precipitation) and parameters (i.e. network intake flow, daily time window, and daily slot period) on the percentages of plots abandoned for irrigation and a Flood affordance. We first changed precipitation forcing values from the year 2005 to 2017 while assigning the model parameters their nominal values: network intake flow is 0.09 m³ s⁻¹ (the river flow is considered as non-limiting), the daily time window is 12 hours and the daily slot period is 10 days. Second, we have set precipitation forcing to the year 2017, and we changed parameter values one at a time (low values of the network intake flow mimics the low-flow period) while assigning the other parameters their nominal values. We repeated each simulation for the two model configurations DailySlots and NoSlots. Simulation runs and results are summarized in the table S2 and the graphic S6 available in the Supplementary Material, respectively.

When varying only the precipitation forcing, the percentage of plots with irrigation abandon is sensitive to precipitation only when the network is coordinated (DailySlots). This is because precipitation constrains the upcoming of Flood affordances on the upstream plots but doesn't influence the downstream plots where irrigation is not possible anyway in the absence of network coordination (NoSlots).

When precipitation forcing does not vary, the percentage of plots with a Flood affordance is sensitive to the network intake flow parameter when the network is not coordinated (NoSlots). No plots are abandoned for irrigation when intake flow exceeds a certain threshold (i.e. 0.35 $m^3 s^{-1}$). This is because the increase in network intake flow parameter increases the percentage of plots with Flood affordances for the downstream plots. The lower the flow at the water intake, the more interesting is the coordination of the network (DailtySlots) to prevent irrigation abandons. In addition, the longer the duration of the daily time window parameter (i.e. the duration the irrigator can act each day) the smaller the percentage of abandoned plots. The longer daily time window parameter facilitates the exploitation of time slots by irrigators, which is the difficulty encountered by irrigator 2 (see Section 3.2.2) who said he can't further increase his daily workload.

Concerning the effect of the modalities of network coordination through daily slots, a period of 8 consecutive days (the two first days for Slot A, the next two days for Slot B, day 5 for Slot C, and the last 3 days for Slot C) prevent abandoned plots at the end of the simulated campaign.

## 4.2 Results about the case study

Model outputs about the case study are consistent with the elements identified during the field surveys. Namely, spatially, the abandonment of the network coordination through daily slots leads to the abandonment of irrigation on the most downstream plots (we observe that in the simulation results through the fact that Flood affordances are absent in downstream plots). The simulated affordances show that the absence of network coordination does not preserve irrigation options on these plots which have the least water supply. Temporally, the absence of network coordination increases the options for the irrigator agents to irrigate. The abandonment of traditional sharing of water releases new temporal leeways on a selected portion of the irrigated command area (the number of Flood affordances in upstream plots are greater in the NoSlots model configuration than in the DailySlots one). This is in line with the ongoing operational changes of individualizing irrigation management that has been described in similar mid-mountain gravity systems (Loubier and Garin, 2013):

- Decrease overtime in the number of farmers, and therefore in the number of workers to maintain the canals. The less the canals are maintained, the fewer plots they serve, or the smaller the flow of water they can carry, because of overflow or because water no longer circulates in certain secondary canals. There is, therefore, less interest in coordinating the network under these conditions, and some farmers have gradually lost interest, with few members coming to the meetings.
- The aging of the technicians to do a minimum of maneuvers at the network junctions and floodgates, and the loss of their knowledge when they have no successor.
- Daily slots are not very flexible, and some farmers prefer to be able to irrigate the plots well supplied with water when they want to, leaving out others that are not well-supplied anyway and that they consider unprofitable.

In addition, we learn in the simulations that the counterpart of the absence of network coordination is an increase in the number of internal conflicts among irrigators (materialized in our study by an increase in the number of AskMoreWater affordances as explained in Section 3.2.1.). Moreover, the simulations highlight the consequences of the heterogeneity of the irrigators' interests within the irrigation scheme according to the location and number of their plots. Irrigator 2 is more penalized by the temporal constraint of traditional sharing of water than irrigator 4 who has an interest in preserving the whole spatial area served by the network. It seems that the choices irrigator 2 makes upstream will make observable differences for all irrigators downstream, which does not appear to be the case for irrigator 4. Simulations point out that it is difficult to implement dynamic irrigation management in this type of ancestral collective institution and maintain suitable water provisioning all across the network.

Lastly, the sensitivity analysis of the model allows the identification of intermediate modalities of network coordination not mentioned by the irrigators. Indeed, coordination through daily slots reduced to a period of 8 days instead of 10 seems optimal (see Section 4.1). In addition, the sensitivity analysis makes it possible to identify the hydro-climatic conditions in which the coordination of the network is advantageous to preserve irrigation as much as possible. An intermediate arrangement could be starting the irrigation campaign without network coordination (i.e. not collectively binding), but considering the implementation of network coordination over 8 days if the flow entering the network from the river is too low (i.e. less than 0.35 $m^3 s^{-1}$).

## 4.3 Benefits and limitations of the use of affordances in ABM as intermediate outcomes

Output interpretation remains highly complex for a large number of ABMs (Lee et al., 2015). Cause-effect relationships cannot always be easily interpreted from the model outputs when agent behavior quickly adapts to a changing environment (Letcher et al., 2013), such as erratic water resources when irrigating. Simulating explicitly the options of the agents through affordances and analyzing the population dynamics of affordances as an additional output of an ABM has benefits to (1) help characterize interaction situations between agents and their environment, and (2) provide insight into what the model does and how the agents are represented in the model.

First, Ferber (1999) has described several interaction situations between agents depending on their goals, capacities, and resources. The abandonment of network coordination leads to what Ferber defines as an overload situation: irrigator agents have compatible goals, sufficient capacities, and insufficient water resources for supplying the whole spatial area. The increase in the number of AskMoreWater affordances following the abandonment of traditional sharing of water indicates an increase in overload situations. We are moving from an individual management problem materialized in the model by the simultaneous presence of Flood affordances within the group of plots of a single irrigator agent who fails to irrigate all of his plots due to lack of time (e.g. case of irrigator agent 4 in DailySlots model configuration) to a collective management problem materialized by the simultaneous presence of Flood affordances spread among plots belonging to several irrigator agents but situated along the same branch of the network. This increases the number of AskMoreWater affordances that reveals a lack of water availability in the canal (e.g. the case of irrigator agent 2 in NoSlots model configuration). The triggering of Flood actions are then no longer dictated primarily by the daily time window parameter (i.e. the duration the irrigator can act each day) but by the network intake flow parameter and precipitation forcing (see Section 4.1). The nature and the dynamic of the affordances, therefore, help characterize the cause of such overload situations. The need for some irrigators to diminish their temporal constraint to control irrigation more finely is likely to create a vicious circle by transferring personal constraints onto the collective functioning by abandoning network coordination. The collective institution then becomes less able to absorb periods of low-flow/irregular precipitation, thus diminishing the interest that irrigators have in it and encouraging the ongoing process of individualization of irrigation.

Second, joint sensitivity analysis of final and intermediate model outcomes helps detect some threshold effects of the model. For instance, no plots are abandoned for irrigation when the network intake flow parameter exceeds a certain threshold as all the downstream plots contain Flood affordances. It is also useful to help reveal biased indicators. In the DailySlots configuration, plot abandonment for irrigation increases with the number of rainy days. This is due to a decrease in the number of Flood affordances which are not generated in case of precipitation, in line with what irrigators say about not irrigating during rainy days. The decrease in the number of Flood affordances indirectly increases the number of days spent without irrigation on the plots, from which the indicator of plot abandonment is computed. It is a bias of the indicator of plot abandonment for the DailySlots configuration during rainy years. This is not the case of the year 2017 (82 rainy days from May to September) used as a reference in our study which is the second least rainy year after 2015 over years 2005-2017 used in the analysis (94 rainy days on average). Rather than using the indicator of plot abandonment, it would be better to directly simulate the consequences of the agent operations into agronomic impacts for plants that would allow better assessment of their biophysical consequences. A good candidate could be the Water Stress Index (Jones, 1992) which controls crop growth and results from both irrigation and precipitation.

Furthermore, the use of affordances limits the capacity of agents in terms of anticipation by making them repeat the same action several times in a row. This is not surprising as our mobilization of the affordance concept inherits from Gibson's ecological approach in which the actor's perception of his possible actions is direct, meaning the capture of information occurs during the action (Luyat & Regia-Corte, 2009). Agent trajectories are thus mainly based on a perception-action loop in line with Afoutni (2015) and Guerrin et al. (2016), without calling for complex decision algorithms. However, human agents and communities actively re-evaluate their beliefs, values, and functioning (Filatova et al., 2013). A perspective of this work is therefore to extend the approach to a perception-cognition-action loop, using for instance elaborated Belief-Desire-Intention algorithms, as done for operational decision-making by Martin-Clouaire (2017) and also to include behavioral norms. It might take advantage of the affordances of the agents to directly nourish their beliefs.

## 4.4 Generality and upscaling perspectives

The conceptual structure of the WatASit model is designed to be as generic as possible. To apply it to other gravity networks, we should collect information about each farm spatial entity (i.e. number of irrigators per farm, maximum simultaneous irrigation operations per irrigator, duration, daily time window, location of irrigated plots, decision-making rules) and each irrigation network (i.e. network intake flow, piloting rules, branch junctions, network seepages, floodgate functioning, list of the plots served by each network branch, and institutional arrangement among irrigators). New spatial entities (i.e. farm plots of each irrigator which are part of the irrigation scheme) could be initialized from GIS data that are now widely available (e.g. the Registre Parcellaire Graphique database in France). In our case study, the interdependencies between irrigation and non-irrigation operations were not taken into account because the farmers said they systematically gave priority to irrigation. In other case studies with significant interdependencies, for instance between harvesting and irrigation (Merot et al., 2008), it is possible to add an affordance specific to the harvesting operation. It allows integrating of harvest affordances into the decision-making of the agents and explicitly takes into account their duration in their workload. Applying the model on pressurized or mixed networks would be possible under the condition of modifying the sub-model of the hydraulic network which is currently not adapted. Upscaling the model at the catchment scale including several irrigation networks would require modeling their influences on hydrology, depending on the water access (surface, groundwater). It would allow representing the mutual influences between the upstream and downstream networks, for instance through a coupling between the WatASit model and a distributed hydrological model.

## 5. Conclusions

In this study, we have proposed the situated WatASit ABM based on the concept of affordance (Gibson, 1977). The objective was to use affordances as intermediate outcomes of the ABM to make explicit the change in agents' options due to a change in the institutional arrangement. We illustrated the approach on a typical gravity network in the South-East of France to assess how the abandonment of traditional sharing of water has impacted irrigators' options to irrigate. We simulated a typical dry year irrigation campaign under two institutional arrangements (i.e. traditional coordination through daily slots and its abandonment). Simulation results were consistent with field surveys in terms of abandonment of irrigation. They revealed an increase in the number of internal conflicts among irrigators as the counterpart of the abandonment of traditional sharing of water. They also highlight the consequences of the heterogeneity of the irrigators' interests within the collective institution. The sensitivity analysis allowed identifying optimal coordination modalities and relevant hydro-climatic conditions for each institutional arrangement. The key benefits of using affordances in ABM lied in analyzing the dynamics of the population of affordances. It provided additional outputs allowing the characterization of the interaction between agents and their environment, and provided new insight into what the model does and how the agents are represented in the model. As our agents appear cognitively limited to schedule irrigation in advance, the first follow-up of this work is to provide them with more elaborated decision-making algorithms. The second key follow-up is to conduct a statistical analysis to demonstrate that changes in affordances provided to the agents are a significant predictor of changes in their behavior, and for further assessing how changes in model outcomes result from changes in affordances provided to the agents.

## Acknowledgments

The authors would like to thank the Zone Atelier Bassin du Rhône (ZABR) and Agence de l'Eau Rhône-Méditerranée-Corse for funding the RADHY Buëch project to which this work contributes. They also thank the French Ministry of Higher Education and Research for the funding provided through AgroParisTech, and Météo-France for providing the SAFRAN reanalysis. Finally they thank the anonymous reviewers for their supportive work.